\documentclass[twocolumn,aps,prb,showpacs,amsmath,amssymb]{revtex4}
\usepackage{graphicx,dcolumn,bm}

\begin{document}

\title{Capacitive interaction model for Aharonov-Bohm effects of a quantum Hall antidot}

\author{W.-R. Lee}
\affiliation{School of Physics, Korea Institute for Advanced Study, Seoul 130-722, Korea}
\author{H.-S. Sim}
\affiliation{Department of Physics, Korea Advanced Institute of Science and Technology, Daejeon 305-701, Korea}

\date{\today}

\begin{abstract}

We derive a general capacitive interaction model for an antidot-based interferometer in the integer quantum Hall regime, and study Aharonov-Bohm resonances in a single antidot with multiple bound modes, as a function of the external magnetic field or the gate voltage applied to the antidot. The pattern of Aharonov-Bohm resonances is significantly different from the case of noninteracting electrons. The origin of the difference includes charging effects of excess charges, charge relaxation between the bound modes, the capacitive interaction between the bound modes and the extended edge channels nearby the antidot, and the competition between the single-particle level spacing and the charging energy of the antidot. We analyze the patterns for the case that the number of the bound modes is 2, 3, or 4. The results agree with recent experimental data.

\end{abstract}



\maketitle

\section{Introduction}

An antidot is a potential hill in a two-dimensional electron system. When a strong perpendicular magnetic field $B$ is applied, the system shows the quantum Hall effects, and there appear bound modes of quantum Hall edge states around the antidot. As the modes enclose magnetic flux, they are governed by the Aharonov-Bohm (AB) effect.~\cite{Aharonv59} When the bound modes couple with extended edge channels via tunneling, the antidot shows AB resonance peaks in electron conductance, as a function of $B$ or the gate voltage $V_\mathrm{BG}$ applied to the antidot. An antidot is a useful tool for detecting and studying localized quantum Hall edge states.

There are experimental evidences that electron interactions play an essential role in an antidot in the integer quantum Hall regime.~\cite{Sim08,Ford94,Kataoka00,Kato09a,Maasilta,Kataoka99,Karakurt,Kataoka02,Goldman08,Gould96,Kato09b} In an antidot with one or two bound modes, the interactions cause interesting phenomena such as $h/2e$ AB effects,~\cite{Ford94,Kataoka00,Kato09a} charging effects,~\cite{Kataoka99} and Kondo effects.~\cite{Kataoka02} AB effects in an antidot with modes more than 2,~\cite{Goldman08} spectator behavior in an antidot molecule,~\cite{Gould96} and charge screening effects~\cite{Karakurt,Kato09b} have also been reported.

Some of the experimental results have been theoretically understood. A phenomenological model for an antidot captures Aharonov-Bohm physics of antidot bound modes as well as the capacitive interactions of excess charges around the antidot; other approaches, such as computations with local-density-functional approximation,~\cite{Ihnatsenka06,Ihnatsenka09} will be useful for studying charge screening. This model successfully describes the $h/2e$ AB effects and Kondo effects in an antidot~\cite{Sim03} as well as the spectator behavior in an antidot molecule,~\cite{Lee10} and it agrees with a Hartree-Fock numerical calculation for an antidot.~\cite{Hwang04}

This may motivate one to extend the model to a general antidot-based interferometer in the integer quantum Hall regime. The extension will be useful for exploring generic aspects of an antidot system with multiple modes, and also for studying antidots in the fractional quantum Hall regime. It may be applicable, with modification, to other quantum Hall interferometers.~\cite{Ofek10,Zhang09,Camino05}

In this paper, we develop a capacitive interaction model applicable to a general situation of antidots, and apply it to an antidot with $\nu_c = 2,3,4$, where $\nu_c$ is the local filling factor around the antidot and equals the number of antidot bound modes. The application demonstrates the generic features by charging effects, charge relaxation between bound modes, the interaction between bound modes and the extended edge channels nearby the antidot, and the competition between charging energy and  single-particle level spacing. It provides systematic understanding of the AB effects in electron conductance $G_\mathrm{T}$ through the antidot as a function of $B$ or $V_\textrm{BG}$. For $\nu_c \ge 3$, the AB effects can deviate from the $h/\nu_c e$ AB oscillation in which there appear $\nu_c$ resonance peaks with equal height and equal peak-to-peak spacing within one period of the peaks. Moreover, the case of $\nu_c=3$ shows resonance signals significantly different from that of $\nu_c = 2,4$. The difference comes from the fact that the interaction between the bound modes and the extended edge channels is much stronger in the case of $\nu_c=3$. The predictions agree with recent experimental data.~\cite{Goldman08}

This paper is organized as follows. In Sec.~II, we derive the capacitive interaction model. In Sec.~III, we summarize nontrivial features of experimental data~\cite{Ford94,Kataoka00,Kato09a,Goldman08} in an antidot with $\nu_c = 2,3,4$. In Sec.~IV, we apply the capacitive interaction model to the antidot, and analyze the AB resonance pattern in the regime that the charging energy is much larger than the single-particle level spacing. In Sec.~V, we consider finite single-particle level spacing. Section~VI provides the summary.

\section{Capacitive interaction model}

In this section, we derive the capacitive interaction model applicable to a general situation of an antidot system under a strong magnetic field $B_0 \gg \Delta B$, where $\Delta B$ is the period of AB resonances of the system. For simplicity, the system is considered to be in the tunneling regime that the antidot bound modes are weakly coupled to extended edge channels, and to be in the regime of zero bias and zero temperature; note that we ignore Kondo effects.~\cite{Sim03} With this simplification, we first describe bound modes in the noninteracting limit, and then the interaction between them.

\begin{figure}[t]
\centering\includegraphics[width=0.48\textwidth]{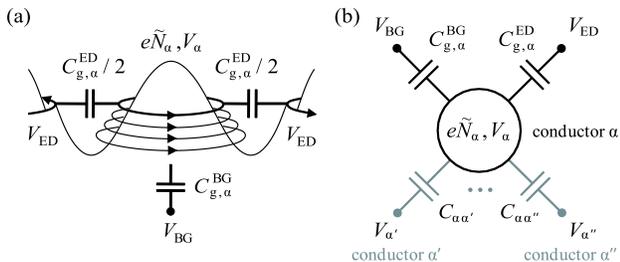}\\
\caption{(a) Schematic view for an antidot under a strong magnetic field $B$. Excess charges $e\tilde{N}_\alpha$ accumulated in mode $\alpha$ interact with those of the other modes (not shown), the extended edge channels nearby the antidot, and the backgate. $V_\alpha$, $V_\mathrm{BG}$, and $V_\mathrm{ED}$ are the voltages of mode $\alpha$, the backgate, and the extended edge channels, respectively. $C_{g, \alpha}^\mathrm{ED}$ ($C_{g, \alpha}^\mathrm{BG}$) is the capacitance between mode $\alpha$ and the extended edge channels (the backgate).
(b) Electric circuit model equivalent to an antidot system with multiple modes. $C_{\alpha \alpha'}$ is the capacitance between modes $\alpha$ and $\alpha'$.}
\label{AD-1-Model}
\end{figure}

\subsection{Bound modes of noninteracting electrons}

We consider an antidot system, which could be a single antidot or multiple antidots, such as an antidot molecule; see a schematic view of an antidot in Fig.~\ref{AD-1-Model}(a). In the integer quantum Hall regime, the bound modes of edge states are formed around the potential of each antidot, which is usually assumed to slowly vary on the scale of magnetic length ($\sim B_0^{-1/2}$). Each mode corresponds to a Landau level whose energy varies to pass the Fermi level near its antidot. For a single antidot, the number of bound modes is identical to the local filling factor $\nu_c$ around the antidot; $\nu_c$ is smaller than the bulk filling factor $\nu_b$. Each mode weakly couples to extended edge channels with the same spin nearby the mode via electron tunneling, and also to the other modes with the same spin. The tunneling strengths depend on the geometry of the system. As we focus on the weak tunneling regime, we ignore the tunneling when we discuss the energy of bound modes; the tunneling will be taken into account in the calculation of $G_\mathrm{T}$.

Each mode $\alpha$ has orbitals $m$ with discrete single-particle energy $\xi_{\alpha m}$. $m$ is the angular momentum quantum number of the orbital or the number of the magnetic flux quanta enclosed by the orbital. $\xi_{\alpha m}$ is determined by antidot potential, Landau level energy, and Zeeman energy. As $B$ increases from a value, say $B_0$, by $\delta B$, each orbital $\alpha m$ spatially shrinks toward the center of its antidot, to keep the same number $m$ of magnetic-flux quanta. This results in the increase of $\xi_{\alpha m}$,
\begin{equation}
\tilde{\xi}_{\alpha m}
= \xi_{\alpha m} + \Delta\xi_{\alpha}\frac{\delta B}{\Delta B_{\alpha}},
\label{EnergyLevel}
\end{equation}
where $\Delta \xi_\alpha$ is the single-particle level spacing of mode $\alpha$
and $\Delta B_{\alpha}$ is the period of AB resonances by mode $\alpha$.
The linear dependence of $\tilde{\xi}_{\alpha m}$ on $\delta B$ is valid for
$\Delta B_\alpha \ll B_0$. In the noninteracting limit, the period is determined by the area $S_\alpha$ enclosed by the single-particle orbital, $\Delta B_\alpha = \phi_0 / S_\alpha$, where $\phi_0$ is the flux quantum.

Equation~\eqref{EnergyLevel} describes AB resonances in the noninteracting limit. As $\delta B$ increases, the energy of each orbital passes, one by one, through the Fermi level. When the energy matches with the Fermi level, electrons tunnel in and out of the orbital, showing a resonance peak in $G_\mathrm{T}$.

In the case that there are multiple modes in the system, each mode results in its own AB resonance peaks independently of the others in the noninteracting limit. The peak height and width are determined by the tunneling strength of the mode to extended edge channels.

\subsection{Coulomb interactions between bound modes}

We turn on electron interactions. In the integer quantum Hall regime, they may be well described~\cite{Sim08,Sim03,Hwang04,Lee10} by capacitive interactions between excess charges accumulated in bound modes, and with those in the extended edge channels nearby the antidot system.

Excess charges accumulated in each mode depend on $\delta B$. As $\delta B$ increases, each orbital of a mode moves toward the center of its antidot to keep the same magnetic flux quanta enclosed by it, which results in the shift of the electron density of the mode. The accumulated excess charge $e \tilde{N}_\alpha$ in mode $\alpha$ due to $\delta B$ is written as
\begin{equation} \label{ElectronNumber}
e \tilde{N}_{\alpha} = e \bigg(N_{\alpha} + \frac{\delta B}{\Delta B_{\alpha}}\bigg),
\end{equation}
where $e~(< 0)$ is the unit of electron charge, and $N_\alpha$ is the total number of electrons occupying mode $\alpha$ at $B_0$. $1/\Delta B_\alpha$ gives the rate of the accumulation. $\Delta B_\alpha$ is a parameter of our model, and it is the period of AB resonances. In the presence of the interaction, $\Delta B_\alpha$ is different from $\phi_0 / S_\alpha$, therefore, $\phi_0 / \Delta B_\alpha$ is interpreted as the {\em effective} area enclosed by mode $\alpha$.

We treat the effect of the capacitive interactions between excess charges, by generalizing an electric circuit model used for a quantum dot.~\cite{Wiel02} Figure~\ref{AD-1-Model}(b) shows the circuit equivalent to the antidot system, where the excess charges of bound modes, the backgate, and extended edge channels interact with each other capacitively; the capacitive interaction between the external voltage sources is ignored. The capacitance between mode $\alpha$ and the backgate (the extended edge channels) is denoted by $C_{g,\alpha}^\mathrm{BG}$ ($C_{g,\alpha}^\mathrm{ED}$). The charge $e\tilde{N}_\alpha$ in Eq.~\eqref{ElectronNumber} is related with the voltages (measured from the ground) $V_{\alpha'}$, $V_\mathrm{BG}$, and $V_\mathrm{ED}$ of mode $\alpha'$, the backgate, and extended edge channels as
\begin{eqnarray}
e\tilde{N}_{\alpha}
& = & |C_{g,\alpha}^\mathrm{BG}|(V_{\alpha} - V_\mathrm{BG})
+ |C_{g,\alpha}^\mathrm{ED}|(V_{\alpha} - V_\mathrm{ED})\nonumber\\
&& + \sum_{\alpha'\neq\alpha}|C_{\alpha\alpha'}|(V_{\alpha} - V_{\alpha'}).
\label{CircuitEq}
\end{eqnarray}
This relation is rewritten as
\begin{equation} \label{ExcessCharge}
\delta Q_{\alpha} \equiv e N_{\alpha} + Q_{\alpha}^\mathrm{G} + e \frac{\delta B}{\Delta B_{\alpha}} = \sum_{\alpha'}C_{\alpha\alpha'}V_{\alpha'}.
\end{equation}
Here we have introduced the excess charge $\delta Q_\alpha$ of mode $\alpha$,
total capacitance $C_{\alpha\alpha} = |C_{g,\alpha}| + \sum_{\alpha'\neq\alpha}|C_{\alpha\alpha'}|$, mutual capacitance $C_{\alpha\alpha'} = -|C_{\alpha\alpha'}|~(\alpha\neq\alpha')$, gate capacitance $C_{g,\alpha} = C_{g,\alpha}^\mathrm{BG} + C_{g,\alpha}^\mathrm{ED}$, and gate charge $Q_{\alpha}^\mathrm{G} = |C_{g,\alpha}^\mathrm{BG}|V_\mathrm{BG} + |C_{g,\alpha}^\mathrm{ED}|V_\mathrm{ED}$. The expression of $\delta Q_\alpha$ has the compensation of gate charge  $Q_\alpha^\mathrm{G}$, in addition to the charge accumulation due to $\delta B$ in $e\tilde{N}_\alpha$; see Eq.~\eqref{ElectronNumber}. From Eq.~\eqref{ExcessCharge}, one obtains the charging energy of the antidot system,
\begin{equation} \label{ChargingEnergy}
E_\mathrm{ch}(\{\delta Q_{\alpha}\})
= \frac{1}{2}\sum_{\alpha}V_{\alpha}\delta Q_{\alpha}
= \sum_{\alpha \alpha'}U_{\alpha\alpha'}
\delta Q_{\alpha}\delta Q_{\alpha'}/e^2,
\end{equation}
where $U_{\alpha\alpha'} \equiv e^2(C^{-1})_{\alpha\alpha'}/2$.

By combining the single-particle energy and the charging energy in Eqs.~\eqref{EnergyLevel} and \eqref{ChargingEnergy}, one has the expression of the ground-state energy of the antidot system,
\begin{equation}\label{TotalEnergy}
E(\{\delta Q_{\alpha}\}) = \sum_{\alpha m}\tilde{\xi}_{\alpha m}n_{\alpha m}
+ E_\mathrm{ch}(\{\delta Q_{\alpha}\}),
\end{equation}
where $n_{\alpha m}$ is the occupation of orbital $\alpha m$. The part of $E$ depending on $\delta B$ is $\sum_{\alpha}\big[\Delta\xi_\alpha N_{\alpha} + \sum_{\alpha'} 2U_{\alpha'\alpha}N_{\alpha'}\big]\frac{\delta B}{\Delta B_\alpha}$.

In the regime of $B_0 \gg \Delta B_\alpha$, one can treat the parameters
(such as $C_{\alpha \beta}$ and $\Delta \xi_\alpha$) of Eq.~\eqref{TotalEnergy} as constant over several $\Delta B_\alpha$, as in the constant interaction model.

\subsection{Accumulation and relaxation of excess charges}

In the capacitive interaction model, one studies the change of the ground state
of the system, characterized by the total numbers $\{N_\alpha\}$ of electron occupation in bound modes $\alpha$, as a function of $\delta B$ and $V_\mathrm{BG}$.

$\delta Q_\alpha$ is accumulated continuously as $\delta B$ increases [see Eq.~\eqref{ExcessCharge}], and relaxed when resonant tunneling of electrons is allowed between a mode and an extended edge channel or internally between different modes. The relaxation results in the transition of the ground state, i.e., the change of  $\{ N_\alpha \}$. First, the resonant tunneling of an electron between mode $\alpha$ and an extended edge channel occurs when
\begin{equation} \label{DtunCon}
E(\delta Q_\alpha \pm e, \dots) = E(\delta Q_\alpha, \dots) \pm \epsilon_\mathrm{F},
\end{equation}
showing a resonance peak in conductance $G_\mathrm{T}$; here $\epsilon_\mathrm{F}$ is the Fermi energy. On the other hand, the internal relaxation of an electron between $\alpha$ and $\alpha'$ occurs when
\begin{equation} \label{CotunCon}
E(\delta Q_\alpha \pm e,\delta Q_{\alpha'} \mp e, \dots)
= E(\delta Q_\alpha,\delta Q_{\alpha'}, \dots).
\end{equation}
It occurs via direct tunneling or cotunneling mediated by virtual states.~\cite{note} It does not cause any resonance peak in $G_\mathrm{T}$, but involves internal charge redistribution of the ground state; some cotunneling processes can slightly modify $G_\mathrm{T}$, but we may neglect the modification. In general, the relaxation of multiple electrons can also occur, depending on systems.~\cite{Lee10} The condition for the relaxation of multiple electrons is obtained by combining Eqs.~\eqref{DtunCon} and \eqref{CotunCon}.

It is useful to draw a charge stability diagram~\cite{Wiel02} to study the transition of the ground state; see below.

\begin{figure}[b]
\centering\includegraphics[width=0.48\textwidth]{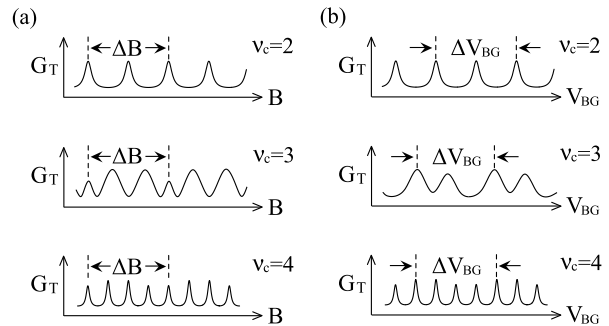}\\
\caption{Schematic views of experimental data for an antidot. (a) Magnetic-field $B$ and (b) back-gate voltage $V_\mathrm{BG}$ dependence of electron conductance $G_\mathrm{T}$ through the antidot. For $\nu_c = 2$, the data show the $h/2e$ AB oscillation.~\cite{Ford94,Kataoka00,Goldman08} For $\nu_c = 3,4$, they deviate from the $h / \nu_c e$ oscillation.~\cite{Goldman08}}
\label{AD-2-Data}
\end{figure}

\section{Experimental data of an antidot}

In this section, we summarize the experimental data of $G_\mathrm{T}$ for an antidot with $\nu_c = 2,3,4$.~\cite{Ford94,Kataoka00,Kato09a,Goldman08} The data disagree with the features of noninteracting electrons discussed in Sec.~II.A. And, the data of $\nu_c = 3,4$ show different features from that of $\nu_c = 2$.

We first mention the dependence of $G_\mathrm{T}$ on $B$. For $\nu_c = 2$, it typically shows the $h/2e$ AB oscillation~\cite{Ford94,Kataoka00,Kato09a} that the two peaks within one AB period are almost identical and have equal peak-to-peak spacing; see the top panel of Fig.~\ref{AD-2-Data}(a). It looks like the periodic structure of one peak per half magnetic flux quantum, instead of two independent peaks (with different height and spacing) per one flux quantum, the expectation from noninteracting electrons.

For $\nu_c = 3,4$, the features of the data deviate from the $h/\nu_c e$ oscillation, an extension of the $h/2e$ AB oscillation.~\cite{Goldman08} For $\nu_c = 3$, the three peaks within one AB period are not identical. Two of them have almost the same height, but are higher than the third [see the middle panel of Fig.~\ref{AD-2-Data}(a)]. For $\nu_c = 4$, two peaks among the four within one AB period are almost identical, but different from the other two (the bottom panel).

Moreover, the dependence of $G_\mathrm{T}$ on $V_\textrm{BG}$ is also nontrivial.~\cite{Goldman08} For $\nu_c = 2, 4$, it resembles the dependence on $B$. In contrast, for $\nu_c = 3$, $G_\mathrm{T}(V_\textrm{BG})$ shows only two peaks ({\em not} three) within one period, i.e., two alternating peak separations within one period; see the middle panel of Fig.~\ref{AD-2-Data}(b). This cannot be understood by simple modification of the $h/\nu_c e$ oscillation.

\section{Single antidot}

In this section, we analyze the features of an antidot with $\nu_c = 2,3,4$, by using the capacitive interaction model. We study the evolution of the ground state as a function of $B$ or $V_\textrm{BG}$.  The transition of the ground state is governed by relaxations of excess charges, and gives rise to AB resonances. We draw charge stability diagrams for the analysis, which has been widely used for the studies of a multiple quantum dot.~\cite{Wiel02} Below, we consider the strong interaction limit of $U_{\alpha\alpha} \gg \Delta\xi_\alpha$. The effects due to finite $\Delta\xi_\alpha$ are discussed in the next section.

We note that some features of $\nu_c = 2,3$ discussed in this section have been already mentioned in the literature.~\cite{Sim03,Sim08,Lee10} We here describe them in more detail (for $\nu_c = 2$ we provide a new analysis), and compare the features of the cases of different $\nu_c$.

\subsection{Antidot with $\nu_c = 2$}

\begin{figure}[t]
\centering\includegraphics[width=0.5\textwidth] {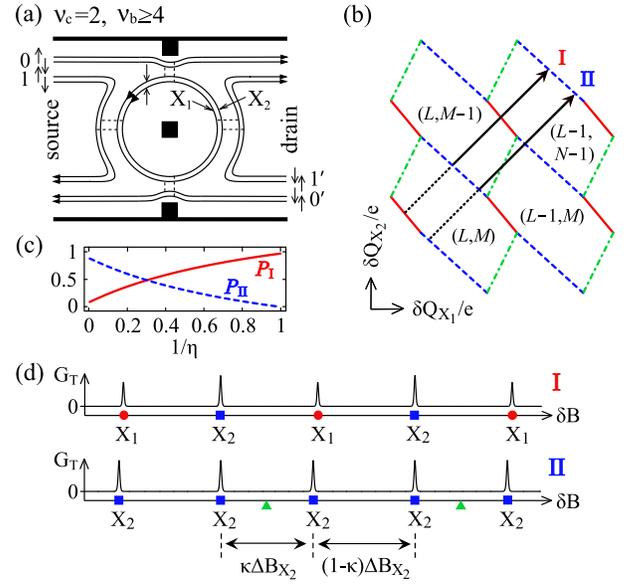}\\
\caption{(Color online) Antidot with $\nu_c = 2$. (a) Antidot modes $X_1$ and $X_2$, edge states (solid line), and electron tunneling (dashed). (b) Charge stability diagram in $(\delta Q_{X_1},\delta Q_{X_2})$ plane. Each of its periodic hexagonal cells represents the ground state of $(N_{X_1},N_{X_2}) = (L,M)$. At blue dashed (red solid) cell boundaries, AB resonances occur via tunneling through $X_2$ ($X_1$), and at green dash-dotted boundaries internal charge relaxations occur between $X_2$ and $X_1$; see  Eqs.~\eqref{DtunCon} and \eqref{CotunCon}. As the magnetic field $\delta B$ increases, $(\delta Q_{X_1},\delta Q_{X_2})$ evolves along a line (solid arrow) of slope $\Delta B_{X_1}/ \Delta B_{X_2}$. Depending on ``initial'' values of $\delta Q_\alpha$ at given field $B_0$, the evolution shows one of two sequences of AB resonances, ``$X_2$-$X_1$" (type I) and ``$X_2$-$X_2$" (II). (c) Probability $P_J (1/\eta)$ of finding the sequences of type $J = \mathrm{I, II}$. (d) Sequence of AB peaks in $G_\mathrm{T}(\delta B)$ for each type in (b). The modes giving the peaks are shown. Triangles represent internal charge relaxation.
For (b)--(d), we choose $\Delta B_{X_1} = \Delta B_{X_2}$, $C_{g,X_1} = 0.5C_{g,X_2} = 0.1C_{X_1X_2}$, $\Delta \xi_\alpha = 0$; the other parameters are given in the Appendix A.}
\label{AD-3-EvolutionNuTwo}
\end{figure}

Figure~\ref{AD-3-EvolutionNuTwo}(a) shows the geometry of an antidot with $\nu_c = 2$. It has two modes, say $X_1$ (inner mode) and $X_2$ (outer), originating from the two spin states of the lowest Landau level. The spatial separation between $X_1$ and $X_2$ is governed by Zeeman splitting (which is enhanced by exchange interactions), while the separation between the modes and extended edge channels is determined by Landau energy gap. The geometry implies $\Delta B_{X_1} \simeq \Delta B_{X_2}$ and $|C_{X_1X_2}| \gg |C_{g,X_2}| > |C_{g,X_1}|$.

The ground-state evolution of the antidot and the resulting AB resonances is studied by analyzing a stability diagram. For $\nu_c = 2$, the evolution of $\{\delta Q_\alpha\}$ follows two different types (I and II) of sequences of AB resonances, depending on how many times the internal relaxations occur per $\Delta B_{X_2}$; see Fig.~\ref{AD-3-EvolutionNuTwo}(b). In type I of $X_2$-$X_1$, the evolution never encounters the internal relaxation, and AB resonances occur sequentially by $X_2, X_1, X_2, X_1, \dots$. In type II of $X_2$-$X_2$, the evolution passes the internal relaxation once per $\Delta B_{X_2}$, and the AB resonances by $X_1$ disappear and are replaced by those by $X_2$; this effect is called the spectator behavior.~\cite{Lee10}

The internal relaxation is characterized by
\begin{equation}
\eta \equiv \frac{(U_{X_1 X_1} - U_{X_1 X_2})\delta B / \Delta B_{X_1}}
{(U_{X_2 X_2} - U_{X_1 X_2})\delta B / \Delta B_{X_2}}
= \frac{C_{g,X_2}}{C_{g,X_1}}\frac{\Delta B_{X_2}}{\Delta B_{X_1}};
\end{equation}
the ratio of energy gains between $\delta Q_{X_1}$ and $\delta Q_{X_2}$ in the internal relaxation between them; see Eq.~\eqref{CotunCon}. In Fig.~\ref{AD-3-EvolutionNuTwo}(b), $\eta$ equals to the ratio of slopes between the dash-dotted relaxation line and the evolution arrow. Based on the antidot geometry mentioned above, we find $\eta > 1$, therefore the relaxation occurs from $\delta Q_{X_1}$ to $\delta Q_{X_2}$.

The internal relaxation is also governed by the inter-mode interaction strength ($\propto |C_{X_1 X_2}|$). As $|C_{X_1 X_2}|$ and/or $\eta~(>1)$ increase, the dash-dotted line in Fig.~\ref{AD-3-EvolutionNuTwo}(b) becomes longer so that the evolution has more chance to pass the internal relaxation. As a result, more sequences (with different ``initial'' values of $\{ \delta Q_\alpha \}$ at $B_0$) follow type II rather than I. When the interaction strength vanishes or $\eta = 1$, the internal relaxation is suppressed and only type I appears. Noninteracting electrons show only type I. This feature appears in the probability $P_J (1/\eta)$ of finding type $J \in \{ \mathrm{I}, \mathrm{II} \}$ in the ensemble of sequences with different values of $\{ \delta Q_\alpha \}$ at $B_0$; see Fig.~\ref{AD-3-EvolutionNuTwo}(c).

We plot the conductance $G_\mathrm{T} (\delta B)$ through the antidot of the sequential tunneling regime in Fig.~\ref{AD-3-EvolutionNuTwo}(d). It is obtained by the standard master-equation approach~\cite{Beenakker91} (see the Appendix A), which is enough to demonstrate the positions and relative heights of AB peaks. Here we assumed low temperature, $k_\mathrm{B} T \ll U_{\alpha \alpha}$, and the backward-reflection regime, in which mode $\alpha$ couples to edge channel $\beta$ with strength $\gamma_{\alpha-\beta}$ as $\gamma_{X_2 - 0\downarrow} > \gamma_{X_1 - 0\uparrow} > \gamma_{X_1 - 1\uparrow}, \gamma_{X_2 - 1\downarrow}$; a similar regime of sequential tunneling and backward reflection will be considered for $\nu_c = 3,4$.

We describe the features of $G_\mathrm{T}$. In type I, each of $X_1$ and $X_2$ shows one peak per period $\Delta B_{X_2}$. The resulting two peaks in $\Delta B_{X_2}$ have different height, because of different $\gamma_{\alpha-\beta}$'s. The peak height of $X_2$ is larger, since $X_2$ is the outer mode. The separation between the two peaks depends on the initial values of $\{ \delta Q_\alpha \}$ at $B_0$. These features are obvious in the noninteracting limit. As the inter-mode interaction increases, the dependence of peak separation on the initial values is weakened, and all the peak-to-peak separations become uniform.

In type II, the two peaks within $\Delta B_{X_2}$ come from $X_2$, therefore, they show the same shape. Unlike type I, the separation between them is determined by interactions. Regardless of the initial values of $\{ \delta Q_\alpha  \}$ at $B_0$, the separations $\kappa \Delta B_{X_2}$ and $(1 - \kappa) \Delta B_{X_2}$ are determined by $\kappa = U_{X_1 X_2}/(U_{X_1 X_2} + U_{X_2 X_2}) = |C_{X_1 X_2}|/(2|C_{X_1 X_2}| + |C_{g,X_1}|)$. In the strong inter-mode interaction limit of $C_{g, X_1}/C_{X_1 X_2} \to 0$, one has $\kappa \rightarrow 1/2$, and the separation becomes $\Delta B_{X_2} / 2$.  In this limit, $G_\mathrm{T}$ shows the well-known $h/2e$ AB effects, and the total energy in Eq.~\eqref{TotalEnergy} becomes $E \simeq U \delta Q_\textrm{tot}^2 / e^2$, where $\delta Q_\textrm{tot} = \sum_\alpha \delta Q_\alpha = 2 e \delta B / \Delta B_{X_2} + \cdots$.

Experimental data of the $h/2e$ AB effects in Refs.~\cite{Ford94,Kataoka00,Goldman08} can be explained by type II; see the top panels of Figs.~\ref{AD-2-Data}(a) and \ref{AD-2-Data}(b). On the other hand, a recent experiment by Kato {\it et al.}~\cite{Kato09a} reported that, as the magnetic field becomes tilted from the perpendicular direction to the two-dimensional system, type II disappears, and instead type I appears. By tilting the magnetic field, one changes the Zeeman energy, therefore controlling the initial values of $\{\delta Q_\alpha \}$ at $B_0$, leading to the transition between I and II.

\subsection{Antidot with $\nu_c = 4$}

\begin{figure}[t]
\centering\includegraphics[width=0.47\textwidth] {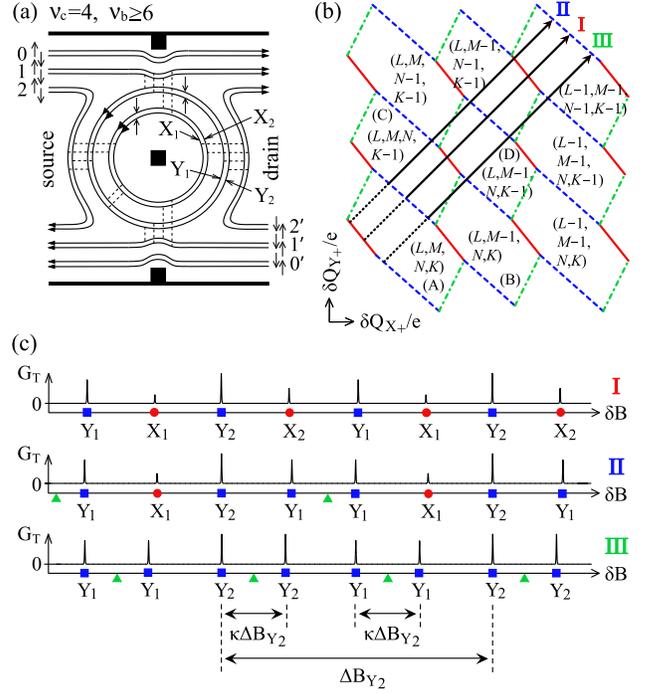}\\
\caption{(Color online) Antidot with $\nu_c = 4$. Each panel corresponds to that of Fig.~\ref{AD-3-EvolutionNuTwo}. (a) Antidot modes $X_1$, $X_2$, $Y_1$, $Y_2$. (b) Charge stability diagram, a periodic structure of four cells (A), (B), (C), (D) in $(\delta Q_{X_+},\delta Q_{Y_+})$ plane. Each cell represents the ground state of $(N_{X_1}, N_{X_2}, N_{Y_1}, N_{Y_2}) = (L, M, N, K)$. As $\delta B$ increases, $(\delta Q_{X_+},\delta Q_{Y_+})$ evolves along a line (solid arrow) of slope $\Delta B_{X_1} / \Delta B_{Y_1}$, while $\delta Q_{X_-}$ and $\delta Q_{Y_-}$ are constant within a cell; one or both of $\delta Q_{X_-}$ and $\delta Q_{Y_-}$ differ by charge $e$ between neighboring cells. The evolution of $(\delta Q_{X_+},\delta Q_{Y_+})$ shows one possible sequence of AB resonances, e.g., ``$Y_2$-$X_2$-$Y_1$-$X_1$" (type I), ``$Y_2$-$Y_1$-$Y_1$-$X_1$" (II), ``$Y_2$-$Y_2$-$Y_1$-$Y_1$" (III). (c) Sequence of AB peaks in $G_\mathrm{T}(\delta B)$ for each type in (b). For (b and c), we choose $\Delta \xi_\alpha = 0$, $\Delta B_{X_1} = \Delta B_{Y_1}$, $C_{g, X_1} = 0.5C_{g, Y_1} = 0.25C_{X_1 Y_1}$, $C_{X_1 X_2} = C_{Y_1 Y_2} = 8 C_{X_1 Y_1}$; the other parameters are in the Appendix A.}
\label{AD-4-EvolutionNuFour}
\end{figure}

We consider an antidot with $\nu_c = 4$. The geometry is shown in Fig.~\ref{AD-4-EvolutionNuFour}(a). It has four modes, $X_1$, $X_2$, $Y_1$, and $Y_2$ (from the innermost one). $X_1$ and $X_2$ ($Y_1$ and $Y_2$) originate from the two spin states of the lowest (second) Landau level. The spatial separation between $X_1$ and $X_2$ and that between $Y_1$ and $Y_2$ are determined by the Zeeman gap, while the other separation (e.g., between $X_1$ and $Y_1$) is governed by the Landau gap. For $\nu_c \ge 3$, the exchange enhancement of the Zeeman gap is much smaller than the Landau gap.~\cite{Xu95} This implies $|C_{X_1 X_2}|, |C_{Y_1 Y_2}| \gg |C_{X_1 Y_1}| > |C_{g, Y_1}| > |C_{g, X_1}|$. Moreover, $X_1$ and $X_2$ ($Y_1$ and $Y_2$) can be treated almost equally so that they have the same values of $\Delta B_\alpha$, $\Delta \xi_\alpha$, $U_{\alpha \alpha}$, $U_{\alpha Y_{1(2)}}$($U_{\alpha X_{1(2)}}$), and $C_{g, \alpha}$. For simplicity, we further approximate $\Delta B_{X_1} \simeq \Delta B_{Y_1}$.

The symmetry between $X_1$ and $X_2$ ($Y_1$ and $Y_2$) simplifies the stability diagram. We introduce new definitions of $\delta Q_{X_\pm} = \delta Q_{X_1} \pm \delta Q_{X_2}$ and $\delta Q_{Y_\pm} = \delta Q_{Y_1} \pm \delta Q_{Y_2}$, and notice that the dependence of $\delta Q_{X_-}$ and $\delta Q_{Y_-}$ on $\delta B$ is negligible for $\Delta B_{X_1} \simeq \Delta B_{X_2}$ and $\Delta B_{Y_1} \simeq \Delta B_{Y_2}$, which is valid for a usual antidot with $S_{X_1}, S_{Y_1} \gg |S_{X_1} - S_{X_2}|, |S_{Y_1} - S_{Y_2}|$. In this case, it is enough to draw the diagram in the subspace $(\delta Q_{X_+}, \delta Q_{Y_+})$ instead of the full space $(\delta Q_{X_1}, \delta Q_{X_2}, \delta Q_{Y_1}, \delta Q_{Y_2})$. $\delta Q_{X_-}$ and $\delta Q_{Y_-}$ are constant within a given cell of the diagram [see Fig.~\ref{AD-4-EvolutionNuFour}(b)] and vary only at cell boundaries by an integer multiple of charge $e$. In terms of $\delta Q_{X_\pm}$ and $\delta Q_{Y_\pm}$, Eq.~\eqref{ChargingEnergy} is rewritten as $E_\mathrm{ch} = \sum_{\mu \mu'}U_{\mu \mu'}\delta Q_\mu \delta Q_{\mu'}/e^2 + \sum_\zeta U_{\zeta\zeta}\delta Q_\zeta^2/e^2$, where $\mu, \mu' \in \{ X_+, Y_+ \}$, $\zeta \in \{ X_-, Y_- \}$, $U_{X_\pm X_\pm} = (U_{X_1 X_1} \pm U_{X_1 X_2})/2$, $U_{Y_\pm Y_\pm} = (U_{Y_1 Y_1} \pm U_{Y_1 Y_2})/2$, and $U_{X_+ Y_+} = U_{Y_+ X_+} = U_{X_1Y_1}$. $X_-$ and $Y_-$ are capacitively decoupled from the others as shown in $E_\mathrm{ch}$.

By analyzing the stability diagram, we find that there are three types, I, II, III, of resonance sequence for $\nu_c = 4$. As in the case of $\nu_c = 2$, the types are characterized by how many times the internal relaxations occur per $\Delta B_{Y_2}$. The evolution passes the internal relaxation never, once, and twice in type I, II, and III, respectively [Fig.~\ref{AD-4-EvolutionNuFour}(b)].

The internal relaxations occur from $X_{1,2}$ to $Y_{1,2}$. In a similar way to $\nu_c =2$, it is understood by the ratio $\eta$ of energy gains between $\delta Q_{X_+}$ and $\delta Q_{Y_+}$ at the relaxation events,
\begin{eqnarray}
\eta \equiv \frac{(U_{X_1 X_1} + U_{X_1 X_2} - 2 U_{X_1 Y_1})/ \Delta B_{X_1}}{(U_{Y_1 Y_1} + U_{Y_1 Y_2} - 2 U_{X_1 Y_1}) / \Delta B_{Y_1}} = \frac{C_{g, Y_1}\Delta B_{Y_1}}{C_{g, X_1}\Delta B_{X_1}}.
\nonumber
\end{eqnarray}
The geometry of edge states indicates $\eta > 1$, which explains the relaxation direction from $X_{1,2}$ to $Y_{1,2}$. We note that the internal relaxation [see Eq.~\eqref{CotunCon}] is forbidden between modes $X_1$ and $X_2$ and between $Y_1$ and $Y_2$ in the symmetric case of $\Delta B_{X_1} = \Delta B_{X_2}$ and $\Delta B_{Y_1} = \Delta B_{Y_2}$.

In Fig.~\ref{AD-4-EvolutionNuFour}(c), we plot $G_\mathrm{T} (\delta B)$ in the sequential tunneling regime. In type I, each of $X_1$, $X_2$, $Y_1$, and $Y_2$ shows one peak per period $\Delta B_{Y_2}$. The resulting four peaks in $\Delta B_{Y_2}$ have different height in general, because of different coupling strengths to extended edge channels. The separation between the peaks by different modes depends on the initial values of $\{ \delta Q_\alpha \}$. As the inter-mode interaction increases, the peak separation becomes independent of the initial values of $\{ \delta Q_\alpha \}$, leading to the uniform peak-to-peak separations.

In type II, two consecutive peaks of the four peaks in one AB period show the same shape. The two peaks come from the same mode $Y_1$ or $Y_2$, and the separation $\kappa \Delta B_{Y_2}$ between the two peaks depends on the interactions as $\kappa = U_{X_1 Y_1}/(2 U_{X_1 Y_1} + U_{Y_1 Y_2} + U_{Y_1 Y_1}) = |C_{X_1 Y_1}|/(4 |C_{X_1 Y_1}| + |C_{g,X_1}|)$, regardless of the initial values of $\{ \delta Q_\alpha  \}$ at $B_0$. In the strong inter-mode interaction limit of $C_{g, X_1}/C_{X_1 Y_1} \to 0$, one has $\kappa \to 1/4$. This behavior results from the internal relaxation. The position of the other two peaks depends on the initial values of $\{ \delta Q_\alpha \}$.

In type III, there occur two consecutive identical peaks by $Y_1$ and the other two identical peaks by $Y_2$ within one AB period. There can also occur another AB resonance sequence such as $Y_2, Y_2, Y_2, Y_1, Y_2, Y_2, Y_2, Y_1, \dots$ but with a small chance. The peak-to-peak separation between the identical peaks is determined by $\kappa \Delta B_{Y_2}$ as in type II. For strong inter-mode interaction, it becomes $\Delta B_{Y_2} / 4$. The separation between the peaks by $Y_1$ and $Y_2$ depends on the initial values of $\{ \delta Q_\alpha \}$ for weak inter-mode interaction, and becomes independent of them, approaching $\Delta B_{Y_2}/4$ as the interaction  increases. Type III is different from the $h/(4e)$ AB effects that all the four peaks have the same shape.

Experimental data for $\nu_c = 4$ in Ref.~\cite{Goldman08} [see the bottom panels of Figs.~\ref{AD-2-Data}(a) and \ref{AD-2-Data}(b)] show two different pairs of identical peaks within one AB period. They can be explained by type III such that one pair occurs by $Y_1$ and the other by $Y_2$. This indicates that electron interactions play a role in the antidot. Note that we do not exclude the possibility of type I of $Y_2$-$Y_1$-$X_2$-$X_1$ which could also show two different pairs of identical peaks accidently.

\begin{figure}[t]
\centering\includegraphics[width=0.5\textwidth] {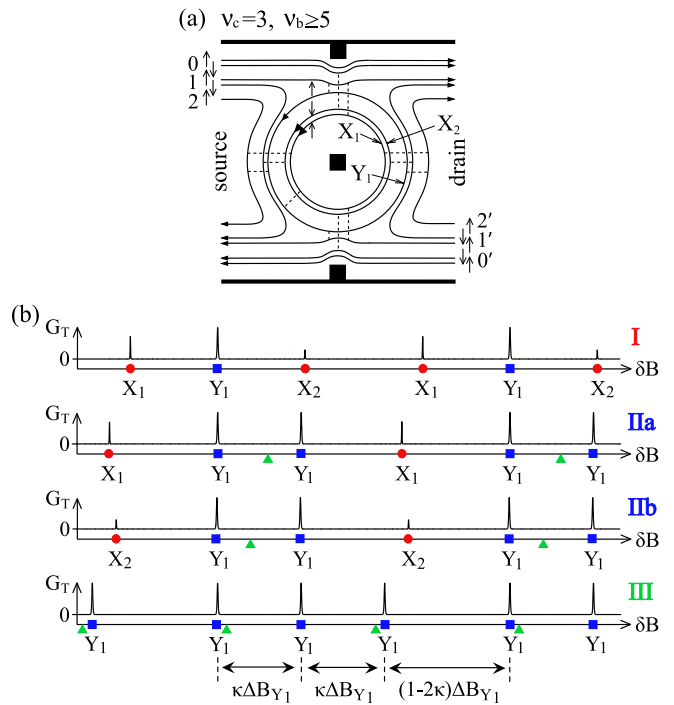}\\
\caption{(Color online) Antidot with $\nu_c = 3$. (a) Antidot modes $X_1$, $X_2$, $Y_1$; see Fig.~\ref{AD-3-EvolutionNuTwo}(a) for details. (b) Possible sequences of AB peaks in $G_\mathrm{T}(\delta B)$. Parameters are in the Appendix A.}
\label{AD-5-EvolutionNuThree}
\end{figure}

\subsection{Antidot with $\nu_c = 3$}

The $\nu_c = 3$ case [see Fig.~\ref{AD-5-EvolutionNuThree}(a)] was previously studied in Ref.~\cite{Lee10}. Here we compare its features with the cases of $\nu_c = 2, 4$.

The antidot has three modes, $X_1$, $X_2$, and $Y_1$ (from the innermost one). The geometry implies $\Delta B_{X_1} \simeq \Delta B_{Y_1}$ and $|C_{X_1 X_2}| > |C_{g, Y_1}| \gg |C_{X_1 Y_1}| > |C_{g, X_1}|$. As in $\nu_c = 4$, we treat $X_1$ and $X_2$ equally so that they have the same values of $\Delta B_\alpha$, $\Delta \xi_\alpha$, $U_{\alpha \alpha}$, $U_{\alpha Y_{1}}$, and $C_{g, \alpha}$. In contrast to $\nu_c = 2,4$,  the separation between modes and extended edge channels  is not governed by the Landau gap but by the Zeeman gap in $\nu_c = 3$. Due to this, AB resonances of $\nu_c = 3$ have different features from $\nu_c =2,4$.

The symmetry between modes $X_1$ and $X_2$ simplifies the stability diagram as in $\nu_c = 4$. By introducing the definition $\delta Q_{X_\pm} = \delta Q_{X_1} \pm \delta Q_{X_2}$, we analyze the ground-state evolution in the subspace of $(\delta Q_{X_+},\delta Q_{Y_1})$. The analysis shows that there are three types I, II, III of resonance sequences, characterized by the number of the internal relaxation events within one AB period [Fig.~\ref{AD-5-EvolutionNuThree}(b)]. The internal relaxations occur from $X_{1,2}$ to $Y_1$.~\cite{Lee10}

We summarize the features of AB peaks. In type I, each mode gives one peak per AB period, showing three different peaks. In type II, two of the three peaks are identical, coming from $Y_1$, and their peak height is larger than the other. In type III, all three peaks are identical, coming from $Y_1$. The separation $\kappa \Delta B_{Y_1}$ between the identical peaks is determined by the interactions as $\kappa = U_{X_1 Y_1}/(2 U_{X_1 Y_1} + U_{Y_1 Y_1}) = |C_{X_1 Y_1}|/(3 |C_{X_1 Y_1}| + |C_{g,X_1}|)$. In the strong inter-mode interaction limit of $C_{g, X_1}/C_{X_1 Y_1} \to 0$, $\kappa \to 1/3$. In this limit, type III shows the $h/(3e)$ AB effects. The dependence of $G_\mathrm{T}$ on $\delta B$ experimentally measured in Ref.~\cite{Goldman08} [the middle panels of Fig.~\ref{AD-2-Data}(a)] can be understood by type II; we do not exclude the possibility of type I that two modes among the three accidently give the peaks of almost equal shape.

We remark that the above description with $\Delta \xi_\alpha = 0$ fails to explain $G_\mathrm{T}(V_\textrm{BG})$ found in Ref.~\cite{Goldman08} where only two peaks appear in one AB period of $V_\textrm{BG}$ [see the middle panels of Fig.~\ref{AD-2-Data}(b)]. This behavior shows an obvious contrast to the case of $\nu_c = 2,4$, in which there appear $\nu_c$ peaks per one AB period of $V_\textrm{BG}$. To understand this difference, we need to take the effect of finite single-particle level spacing, which is the subject of the next section.

\section{Single-particle level spacing}

In the previous section, the analysis of AB effects was restricted to the regime of $\Delta \xi_ \alpha \ll U_{\alpha \alpha}$. In a general situation, single-particle level spacing $\Delta \xi_\alpha$ is not ignorable. One expects that interaction effects will become reduced as $\Delta \xi_\alpha$ increases. In this section, we study how finite $\Delta \xi_\alpha$ modifies the interaction effects found in Sec.~IV.

We study the effect of finite $\Delta \xi_\alpha$ by assuming linear dispersion $\xi_{\alpha m} = \xi_{\alpha 0} + (m - 1)\Delta \xi_\alpha$. Then, the first term of the ground-state energy in Eq.~\eqref{TotalEnergy} is written as $\sum_\alpha \Delta \xi_\alpha(N_\alpha + \delta B / \Delta B_\alpha)^2 / 2$. It is quadratic in $N_\alpha$. It is absorbed into the interaction term of Eq.~\eqref{TotalEnergy} so that the total energy has the same form as the interaction term, $E(\{\delta Q_\alpha\}) = \sum_{\alpha\alpha'}\tilde{U}_{\alpha\alpha'}\delta \tilde{Q}_\alpha \delta \tilde{Q}_{\alpha'}$, but with modification
\begin{eqnarray}
\tilde{U}_{\alpha\alpha'} & = & U_{\alpha\alpha'} + \delta_{\alpha\alpha'}\Delta\xi_\alpha / 2, \label{newU} \\
\delta \tilde{Q}_\alpha & = & e N_\alpha + \tilde{Q}_\alpha^\mathrm{G} + e \delta B / \Delta B_\alpha. \label{newQ}
\end{eqnarray}
Here, $\tilde{Q}_\alpha^\mathrm{G}$ satisfies $\sum_{\alpha'}\tilde{U}_{\alpha\alpha'}\tilde{Q}_{\alpha'}^\mathrm{G} = \sum_{\alpha'}U_{\alpha\alpha'}Q_{\alpha'}^\mathrm{G}$.

\begin{figure}[t]
\centering\includegraphics[width=0.47\textwidth]{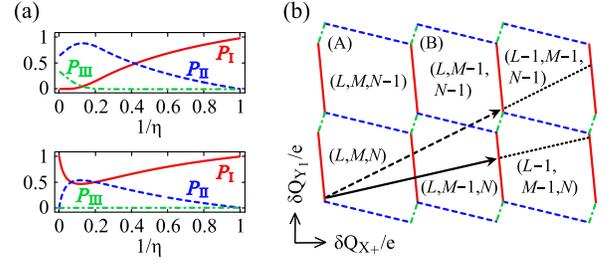}\\
\caption{(Color online) Antidot with $\nu_c = 3$. (a) Probability $P_J(1 / \eta)$ of finding type I, II (IIa +IIb), III (see Fig.~\ref{AD-5-EvolutionNuThree}) for $\Delta \xi_\alpha = 0$ (upper panel) and $\Delta \xi_\alpha = e^2 / C_{X_1X_1}$ (lower). $\eta = {C_{g, Y_1}\Delta B_{Y_1}}/{C_{g, X_1}\Delta B_{X_1}}$. (b) Evolution of $(\delta Q_{X_+},\delta Q_{Y_1})$ as a function of the {\em back-gate voltage} for $\Delta \xi_\alpha = 0$ (dashed arrow) and $\Delta \xi_\alpha = e^2 / C_{X_1X_1}$ (solid). The charge stability diagram has a periodic structure of two cells (A), (B). Each cell represents the ground state of $(N_{X_1}, N_{X_2}, N_{Y_1}) = (L, M, N)$. For (a) and (b), we choose $C_{X_1 Y_1} = 2 C_{g, X_1} = 0.125C_{X_1 X_2}$, $C_{g, Y_1} = 32 C_{g, X_1}$, and $\delta Q_{X_-} = -0.4 e$ for cell (A) and $0.6 e$ for (B).}
\label{AD-6-Backgate}
\end{figure}

We discuss the results of the modification. First, in Eq.~\eqref{newU}, $\Delta \xi_\alpha$ effectively enhances the self-interaction of mode $\alpha$, without modifying intermode interactions. As a result, as $\Delta \xi_\alpha$ increases, the internal relaxations between different modes (therefore type II and III) become suppressed. The suppression is shown for $\nu_c = 3$ in Fig.~\ref{AD-6-Backgate}(a).

Second, $\Delta \xi_\alpha$ also affects $\delta Q_\alpha$ in Eq.~\eqref{newQ}. It should be emphasized that it does not alter the dependence of $\{ \delta Q_\alpha \}$ on $\delta B$. Hence an antidot with finite $\Delta \xi_\alpha$ shows the same dependence of AB peaks on $\delta B$ as the case of $\Delta \xi_\alpha = 0$. However, $\Delta \xi_\alpha$ {\em does} affect the dependence of $\{ \delta Q_\alpha \}$ on $V_\textrm{BG}$. In the stability diagram [see Fig.~\ref{AD-6-Backgate}(b)], the evolution follows along a line of slope $s_{\nu_c} (\Delta \xi_\alpha)$. The expression of $s_{\nu_c}$ is
\begin{equation}\label{VBGslope}
\frac{s_{\nu_c}(\Delta \xi)}{s_{\nu_c}(\Delta \xi = 0)} = \frac{1 + (|C_{g, X_1}| + \nu_c |C_{X_1 \beta}|) \Delta\xi / e^2}{1 + (|C_{g,\beta}| + \nu_c |C_{X_1 \beta}|) \Delta\xi / e^2},
\end{equation}
where $\beta = X_2$ for $\nu_c =2$ and $\beta = Y_1$ for $\nu_c = 3,4$. Here, we have put the same value of $\Delta \xi$ into $\Delta \xi_\alpha$'s.

For a $\nu_c = 3$ antidot with finite $\Delta \xi_\alpha$ comparable with $e^2 / |C_{X_1Y_1}|$, $s_{3}$ can be much smaller than the value of $\Delta \xi_\alpha = 0$. It is because of the geometrical feature [see Fig.~\ref{AD-5-EvolutionNuThree}(a)] that the spatial separation of $Y_1$ from extended edge channels is determined by the Zeeman gap, while that of $X_1$ is determined by the Landau gap. This feature implies $|C_{g,Y_1}| \gg |C_{X_1Y_1}| > |C_{g,X_1}|$. The evolution line with small $s_3$ can pass only the solid cell boundaries in the stability diagram, showing paired peaks by $X_1$ and $X_2$ over several periods; see Fig.~\ref{AD-6-Backgate}(b). Or, depending on the initial value of $\{ Q_\alpha \}$ at $B_0$, it can pass only the dashed cell boundaries, showing paired peaks by $Y_1$. Hence, the sequence of AB peaks shows only two peaks within one period. This behavior may explain the puzzling features of the experimental data~\cite{Goldman08} in the middle panel of Fig.~\ref{AD-2-Data}(b).

In contrast, in the cases of $\nu_c = 2, 4$, there is no drastic change of $s_{\nu_c}$ when $\Delta \xi_\alpha$ changes from 0 to $e^2/|C_{\alpha \beta}|$. Unlike $\nu_c = 3$, these cases have the common geometrical feature that the separation between the outermost mode and edge channels is governed by the Landau splitting, which implies $|C_{X_1X_2}| \gg |C_{g,X_2}| > |C_{g,X_1}|$ for $\nu_c = 2$ and $|C_{X_1Y_1}| > |C_{g,Y_1}| > |C_{g,X_1}|$ for $\nu_c = 4$. With these parameters of capacitances, one notices $s_{\nu_c = 2,4} (e^2/|C_{\alpha \beta}|) \simeq s_{\nu_c = 2,4} (0)$. Therefore $G_\mathrm{T}(V_\textrm{BG})$ shows $\nu_c$ peaks within one AB period, similar to the dependence on $\delta B$.

This finding emphasizes the role of the electron interaction between an antidot and extended edge channels as well as the competition between electron interactions and single-particle physics in an antidot system.

\section{Conclusion}

We have developed a capacitive interaction model for an antidot system in the integer quantum Hall regime, and applied it to an antidot with $\nu_c = 2,3,4$. The predictions of the model agree with the various features of $G_\mathrm{T}(\delta B)$ and $G_\mathrm{T}(V_\textrm{BG})$ observed in experiments. This may support the validity of our model, and show the importance of electron interactions in the antidot.

We summarize the features of an antidot predicted by our model. The common features are (i) charging effects, (ii) internal relaxations from an inner mode to an outer one ($\eta > 1$, independent of $\nu_c$), and (iii) $\nu_c$ resonance peaks in $G_\mathrm{T} (\delta B)$ within one period; in other geometries, internal relaxations can occur in the opposite way.~\cite{Lee10} For larger $\nu_c$, more different types of sequences of the peaks can appear. As the interaction strength increases, the peaks within one period become correlated, i.e., some of them can have the same shape (coming from the same mode) and the peak-to-peak separation becomes equally spaced. For $\nu_c = 3,4$, the peaks show the features deviated from the $h/(\nu_c e)$ oscillation. On the other hand, the feature of $G_\mathrm{T} (V_\mathrm{BG})$ depends on $\nu_c$. For $\nu_c = 2,4$, $G_\mathrm{T} (V_\mathrm{BG})$ shows $\nu_c$ peaks within one period, while for $\nu_c = 3$, it can show only two peaks. This $\nu_c$ dependence comes from the geometrical difference of the spatial separation between the antidot and extended edge channels as well as the competition between the electron interactions and the single-particle level spacing.

In other types of quantum Hall interferometers, such as Fabry-Perot resonators, electron interactions will also play an important role, since the interferometers utilize localized quantum Hall edge states. Some puzzling features (e.g., a checkerboard pattern) of electron conductance through a Fabry-Perot resonator have been experimentally observed.~\cite{Ofek10,Zhang09,Camino05} They may be originated from electron interactions. Some parts of the features can be explained by using a capacitive interaction model which was proposed recently,~\cite{Rosenow07} but the features have not been fully understood yet. To understand them, it will be interesting to extend our model to the Fabry-Perot resonator.

\begin{acknowledgments}

We thank C.J.B. Ford, V.J. Goldman, and M. Kataoka for discussion, and financial support by National Research Foundation (Grant No. 2009-0078437).

\end{acknowledgments}

\appendix

\section{Electron conductance}

We provide the derivation of electron conductance $G_\mathrm{T}$ through the antidot with $\nu_c =2$ in the sequential tunneling regime of $\gamma_{\alpha-\beta} \ll \Delta \xi_\alpha, k_\mathrm{B}T, U_{\alpha\alpha}$. It uses the master-equation approach.~\cite{Beenakker91} The derivation is trivially extended to $\nu_c = 3,4$, so we do not provide the extension.

In the antidot system [see Fig.~\ref{AD-3-EvolutionNuTwo}(a)], electron current $I$ by small bias voltage $V_\mathrm{sd}$ between the source and drain is decomposed into three contributions, $I = I_\mathrm{d} + I_\mathrm{f} - I_\mathrm{b}$. $I_\mathrm{f}$ ($I_\mathrm{b}$) comes from the resonant forward (backward) scattering of an electron from edge channel $1\sigma$ or $0\sigma$ to $1'\sigma$ ($0'\sigma$) via antidot bound modes, while $I_\mathrm{d}$ comes from electrons propagating along the extended edge channels without any scattering by the antidot; $\sigma = \uparrow,\downarrow$ is a spin index. The conductance $G_\mathrm{T}$ can be defined by excluding the trivial contribution of $I_\mathrm{d}$, $G_\mathrm{T} \equiv |I - I_\mathrm{d}|/V_\mathrm{sd} = |I_\mathrm{f} - I_\mathrm{b}|/V_\mathrm{sd}$.

$I_\mathrm{f}$ ($I_\mathrm{b}$) is the net flux of  charge $e$ across the tunneling barrier between the antidot and edge channels $1'\sigma$ ($0'\sigma$),
\begin{equation} \label{Current}
I_{j} = e \sum_{L,M} P_{L,M} \sum_{L',M'} n_{L,M,L',M'} \Gamma_{L,M \rightarrow L',M'}^{\mathrm{d},j},
\end{equation}
where $j = \mathrm{f,b}$. $P_{L,M}$ is the probability of finding the antidot in the ground state of $(N_{X_1},N_{X_2}) = (L,M)$, and $(L,M)$ varies over the ground-state configurations. $\Gamma^{\mathrm{d},j}$ is the rate of the transition from $(L,M)$ to $(L',M')$ via electron tunneling between the antidot and the channel $1'\sigma$ or $0'\sigma$ injected from the drain. The possible values of $(L',M')$ are $(L \pm 1,M)$ and $(L,M \pm 1)$, and $n_{L,M,L',M'} = L+M-L'-M'$ counts the net electrons tunneling from the antidot to the channels $1'\sigma$ and $0'\sigma$ at the transition. $\Gamma^{\mathrm{d},j}$ is expressed as $\hbar \Gamma_{L,M \rightarrow L + 1,M}^{\mathrm{d},j} = \gamma^{\mathrm{d},j}  f(\Delta E_{L,M \rightarrow L + 1,M} + \eta eV_\mathrm{sd})$, $\hbar \Gamma_{L,M \rightarrow L - 1,M}^{\mathrm{d},j} = \gamma^{\mathrm{d},j}~[1 - f(\Delta E_{L - 1,M \rightarrow L,M} + \eta eV_\mathrm{sd})]$, etc., where $\gamma^{\mathrm{d,f}} = \gamma_{X_1 - 1'\uparrow}$, $\gamma^{\mathrm{d,b}} =  \gamma_{X_1 - 0'\uparrow}$, $f(\epsilon) = (1 + e^{\epsilon/(k_\mathrm{B}T)})^{-1}$, $\Delta E_{L,M \rightarrow L',M'}$ is the change of antidot energy at the transition, e.g., $\Delta E_{L,M \rightarrow L + 1,M} = E(L + 1,M) - E(L,M) - \epsilon_\mathrm{F}$, and $0 < \eta < 1$.~\cite{Beenakker91}

The probability $P_{L,M}$ satisfies the detailed balance equations in the stationary situation of $d P_{L,M} / dt = 0$, i.e., $\sum_{L',M'} [P_{L',M'} \Gamma_{L',M' \rightarrow L,M} - P_{L,M} \Gamma_{L,M \rightarrow L',M'}] = 0$. Here, we define the total transition rate $\Gamma \equiv \Gamma^\mathrm{s,f} + \Gamma^\mathrm{s,b} + \Gamma^\mathrm{d,f} + \Gamma^\mathrm{d,b}$, where $\Gamma^{\mathrm{s},j(= \mathrm{f,b})}$ is the transition rate of the antidot ground state via electron tunneling between the antidot and the channel $1\sigma$ or $0\sigma$ injected from the source. In the linear-response regime, the detailed balance equations can be solved by imposing the approximate form $P_{L,M} \simeq P^\mathrm{eq}_{L,M} [1 + w_{L,M} eV_\mathrm{sd}/(k_\mathrm{B}T)]$, where $P^\mathrm{eq}_{L,M} = \mathcal{Z}^{-1} e^{- [E(L,M) - (L + M)\epsilon_\mathrm{F}] / (k_\mathrm{B}T)}$ is the equilibrium distribution, $\mathcal{Z}$ is determined by $\sum_{L,M} P^\mathrm{eq}_{L,M} = 1$, and $w_{L,M}$ satisfies the equation, e.g., $w_{L + 1,M} = w_{L,M} - \eta + (\gamma_{X_1 - 1\uparrow} + \gamma_{X_1 - 0\uparrow})/(\gamma_{X_1 - 1\uparrow} + \gamma_{X_1 - 0\uparrow} + \gamma_{X_1 - 1'\uparrow} + \gamma_{X_1 - 0'\uparrow})$.

Substituting $P_{L,M}$ into Eq.~\eqref{Current}, and totally linearizing Eq.~\eqref{Current} in terms of $V_\mathrm{sd}$, we find the formula,
\begin{eqnarray}
G_\mathrm{T} & = & \frac{e^2}{h}\frac{2 \pi}{k_\mathrm{B}T} \sum_{L,M} \sum_{\alpha = X_1, X_2} P^\mathrm{eq}_{L,M} F_{L,M}^{\alpha}, \label{Conductance} \\
F_{L,M}^{X_1} & = &  \frac{(\gamma_{X_1 - 1\uparrow} + \gamma_{X_1 - 0\uparrow})|\gamma_{X_1 - 1'\uparrow} - \gamma_{X_1 - 0'\uparrow}|}{\gamma_{X_1 - 1\uparrow} + \gamma_{X_1 - 0\uparrow} + \gamma_{X_1 - 1'\uparrow} + \gamma_{X_1 - 0'\uparrow}} \nonumber\\
&& \times f(\Delta E_{L,M \rightarrow L + 1,M}), \nonumber
\end{eqnarray}
\begin{eqnarray}
F_{L,M}^{X_2} & = &  \frac{(\gamma_{X_2 - 1\downarrow} + \gamma_{X_2 - 0\downarrow}) |\gamma_{X_2 - 1'\downarrow} - \gamma_{X_2 - 0'\downarrow}|}
{\gamma_{X_2 - 1\downarrow} + \gamma_{X_2 - 0\downarrow} + \gamma_{X_2 - 1'\downarrow} + \gamma_{X_2 - 0'\downarrow}} \nonumber\\
&& \times f(\Delta E_{L,M \rightarrow L,M + 1}). \nonumber
\end{eqnarray}
We note that Eq.~\eqref{Conductance} is derived in the case where the internal relaxation between antidot modes does not occur around the central position of AB peaks (within peak broadening), thus it does not affect the shape of AB peaks.

Finally, we mention the parameters used in the figures. We choose tunneling strengths $\gamma_{\alpha-\beta}$, assuming that antidot modes are symmetrically coupled to upper and lower edge channels of the system, and that the antidot is in the backward-reflection regime of $I_\textrm{b} > I_\textrm{f}$. In Fig.~\ref{AD-3-EvolutionNuTwo}(d), we choose $k_\mathrm{B}T = 0.02 e^2/C_{X_1 X_1}$, $\gamma_{X_1 - 0\uparrow} = 0.8 \gamma_{X_2 - 0\downarrow}$, and $\gamma_{X_1 - 1\uparrow} = \gamma_{X_2 - 1\downarrow} = 0.1 \gamma_{X_2 - 0\downarrow}$. In Fig.~\ref{AD-4-EvolutionNuFour}(b), $(\delta Q_{X_-}, \delta Q_{Y_-}) = (- 0.2, - 0.3) e$, $(0.8, - 0.3) e$, $(- 0.2, 0.7) e$, $(0.8, 0.7) e$ for (A), (B), (C), (D), respectively. In Fig.~\ref{AD-4-EvolutionNuFour}(c), $k_\mathrm{B}T = 0.02 e^2 / C_{X_1 X_1}$, $\gamma_{Y_1 - 1\uparrow} = 0.85 \gamma_{Y_2 - 1\downarrow}$, $\gamma_{X_2 - 1\downarrow} = 0.6 \gamma_{Y_2 - 1\downarrow}$, $\gamma_{X_1 - 1\uparrow} = 0.4 \gamma_{Y_2 - 1\downarrow}$, $\gamma_{Y_2 - 2\downarrow} = \gamma_{Y_1 - 2\uparrow} = 0.3 \gamma_{Y_2 - 1\downarrow}$, and $\gamma_{X_2 - 2\downarrow} = \gamma_{X_1 - 2\uparrow} = 0.2 \gamma_{Y_2 - 1\downarrow}$. In Fig.~\ref{AD-5-EvolutionNuThree}(b), $\Delta \xi_\alpha = 0$, $\Delta B_{X_1} = \Delta B_{X_2} = \Delta B_{Y_1}$, $C_{g, Y_1} = 8 C_{g, X_1}$, $C_{X_1 Y_1} = 2 C_{g, X_1}$, $C_{X_1 X_2} = 8 C_{X_1 Y_1}$, $k_\mathrm{B}T = 0.005e^2 / C_{X_1 X_1}$, $\gamma_{X_1 - 1\uparrow} = 0.7 \gamma_{Y_1 - 1\uparrow}$, $\gamma_{X_2 - 0\downarrow} = 0.5 \gamma_{Y_1 - 1\uparrow}$, and $\gamma_{X_2 - 1\downarrow} = \gamma_{Y_1 - 2\uparrow} = 0.3 \gamma_{Y_1 - 1\uparrow}$, $\gamma_{X_1 - 2\uparrow} = 0.2 \gamma_{Y_1 - 1\uparrow}$.


\begin{thebibliography}{99}
\bibitem{note} Note that the rate of the cotunneling processes in Eq.~\eqref{CotunCon} can be estimated by Fermi's ``golden rule." For example, the rate of elastic cotunneling processes is obtained as $\Gamma_\textrm{ela} = k_\mathrm{B} T~\Gamma^2 / (2 \pi U^2)$, where $U$ is the antidot charging energy, $\Gamma$ is the broadening of AB resonances, and $k_\mathrm{B} T$ is the thermal energy. For an antidot with $U \sim 10^{-4}$~$e$V, $\Gamma \sim 10^{-5}$ $e$V, and $T \sim 10$~mK, one obtains $\Gamma_\textrm{ela} \sim 1.4 \times 10^{-9}$~$e$V. The time scale of the relaxation via the cotunneling processes (including spin-flip processes) is $\hbar / \Gamma_\textrm{ela} \sim 5 \times 10^{-7}$~sec.
\end{thebibliography}
\end{document}